\documentclass[final,3p,times]{elsarticle}

\usepackage{amssymb}
\usepackage{graphicx,multirow,caption,subcaption,multicol,setspace,amsmath,wrapfig}
\usepackage{commath}
\usepackage[linesnumbered,ruled]{algorithm2e}
\usepackage{algorithmic}
\usepackage{adjustbox}
\usepackage{booktabs,caption}
\usepackage[flushleft]{threeparttable}
\usepackage{lineno}
\usepackage{enumitem}
\setlist{nosep}
\usepackage{rotating,caption}
\usepackage[table]{xcolor}
\usepackage{soul,xcolor}
\usepackage{amsfonts}
\usepackage{makecell,eqnarray}

\usepackage{tikz}
\usetikzlibrary{shapes,arrows,decorations.markings}
\usepackage[EULERGREEK]{sansmath}
\usepackage{floatrow}
\DeclareFloatFont{henry}{\fontfamily{ccr}\fontseries{m}}
\usepackage{float}
\floatstyle{plaintop}
\restylefloat{table}

\biboptions{numbers,sort&compress}

\DeclareMathOperator*{\argmin}{arg\,min}

\usetikzlibrary{shapes}
\usetikzlibrary{shapes.multipart}
\definecolor{CC1}{rgb}{0.09, 0.45, 0.27}
\definecolor{CC2}{rgb}{0.9,0.45,0.1}
\usetikzlibrary{chains,decorations.pathreplacing}
\usetikzlibrary{fit,positioning,shapes.geometric}
\usepackage{tikz}
\usetikzlibrary{shapes,arrows,decorations.markings}
\usetikzlibrary{arrows,calc}
\usetikzlibrary {arrows.meta} 
\usepackage{relsize}

\graphicspath{ {Images/} }

\makeatletter
\newcommand{\algorithmfootnote}[2][\footnotesize]{%
  \let\old@algocf@finish\@algocf@finish
  \def\@algocf@finish{\old@algocf@finish
    \leavevmode\rlap{\begin{minipage}{\linewidth}
    #1#2
    \end{minipage}}%
  }%
}

\journal{Neurocomputing}

\begin{document}
\begin{frontmatter}

\title{Evolution of Neural Architectures for Financial Forecasting: A Note on Data Incompatibility during Crisis Periods}

\author[label5]{Faizal Hafiz\corref{cor1}}
\address[label5]{SKEMA Business School, Université Côte d’Azur, Sophia Antipolis, France}
\ead{faizal.hafiz@skema.edu}
\cortext[cor1]{Corresponding author}

\author[label5]{Jan Broekaert}

\author[label1]{Akshya Swain}
\address[label1]{Department of Electrical, Computer \& Software Engineering, The University of Auckland, New Zealand}

\begin{abstract}

This note focuses on the optimization of neural architectures for stock index movement forecasting following a major market disruption or crisis. Given that such crises may introduce a shift in market dynamics, this study aims to investigate whether the training data from market dynamics prior to the crisis are compatible with the data during the crisis period. To this end, two distinct learning environments are designed to evaluate and reconcile the effects of possibly different market dynamics. These environments differ principally based on the role assigned to the pre-crisis data. In both environments, a set of non-dominated architectures are identified to satisfy the multi-criteria co-evolution problem, which simultaneously addresses the selection issues related to features and hidden layer topology. To test the hypothesis of pre-crisis data incompatibility,  the day-ahead movement prediction of the NASDAQ index is considered during two recent and major market disruptions; the 2008 financial crisis and the COVID-19 pandemic. The results of a detailed comparative evaluation convincingly support the incompatibility hypothesis and highlight the need to select re-training windows carefully.

\end{abstract}

\begin{keyword}
Co-evolution \sep Feature Selection \sep Financial Crisis \sep Financial Forecasting \sep Multi-criteria Decision Making \sep Neural Architecture Search
\end{keyword}

\end{frontmatter}

\section{Introduction}

A major disruption in stock markets arising from a financial event, like the 2007-2008 subprime mortgage crisis, or a global event, like the COVID-19 pandemic, is often associated with a shift in the market dynamics. In particular, a change in market \emph{efficiency} is observed following the 2008 crisis in~\citep{Horta:Lagoa:2014,Anagnostidis:Varsakelis:2016,Smith:2012,Choi:2021}. Similarly, empirical investigations in~\citep{ChandraEtAl2021,Choi:2021,Hafiz:Broekaert:2021,Hafiz:Broekaert:2023} highlight possible inconsistencies in market data prior to and after the breakout of the COVID-19 pandemic. Such disparity in market behaviors represents a learning dilemma in the sense that while the pre-crisis data may deteriorate forecasting performance during the post-crisis period, it is usually not desirable to completely disregard any partially congruous information contained in an otherwise conflicting pre-crisis dataset. The stock market crises, thus, can be thought of as real-life benchmarks to investigate the performance of forecasting models over conflicting data time windows. This note, therefore, aims to investigate and reconcile potential shifts in market dynamics ensuing a significant market disruption or crisis. 


In particular, the goal is to demonstrate how an adapted learning scheme can reconcile aspects of the dynamical shift in stock markets following a major financial crisis. To this end, we examine the potential impacts of incongruous market dynamics on neural architecture optimization, which involves a careful selection of features and neural topology. While our previous investigations in~\citep{Hafiz:Broekaert:2021,Hafiz:Broekaert:2023} developed an extensive multi-criteria \emph{co-evolution} framework for neural architecture optimization, the implications of change in market behaviors were not investigated in detail. For instance, in~\citep{Hafiz:Broekaert:2021}, the impact of pre-crisis data on a co-evolution of neural architecture with scalarized goal were compared through different learning scenarios by considering market data prior to and within the COVID-19 pandemic. The results of this empirical investigation highlighted that pre-crisis data could be inconsistent to a certain degree and can adversely affect the optimization of neural architectures. This pre-crisis data incompatibility investigation was, however, limited to a scalarized goal optimization of neural architectures, which requires \emph{a priori} preference specification from a decision maker to focus on a particular region of the multi-criteria Pareto front. In contrast, the multi-objective co-evolution framework in~\citep{Hafiz:Broekaert:2023} focuses on identifying the entire set of non-dominated architectures. While this framework includes a mechanism to address the data incongruence, the impacts of pre-crisis data on the set of identified architectures were not compared through different learning scenarios. Nevertheless, the results of these earlier investigations underline the possibilities of incompatibility in pre-crisis data, which serves as the motivation for this note. 

 Our primary objective is to empirically test the hypothesis that the information in the pre-crisis market dataset can be incongruent with the altered post-crisis market dynamics. Next, we focus on the reconciliation of market behaviors, assuming the hypothesis to be true. This is accomplished by optimizing neural architectures in two distinct learning environments which assign a different role to the pre-crisis dataset. The first environment assumes that the incompatibility hypothesis is false and includes the pre-crisis dataset for supervised learning, which serves as the baseline. In contrast, the second learning environment augments the co-evolution of neural architecture to use the pre-crisis dataset in a limited capacity. A detailed evaluation setup is designed to compare these learning environments for a day-ahead movement forecasting of the NASDAQ index during two major market disruptions, \textit{i.e.}, the 2008 crisis and the COVID-19 pandemic. The set of non-dominated architectures is identified in a total of eight distinct search scenarios, which represent a particular combination of two market disruptions, two learning environments and two multi-objective search philosophies.

Due to the brevity this note, a detailed discussion on several related issues is omitted; we refer to the following investigations and the references therein for technical indicator-based forecasting~\citep{Bustos:2020}, neural architecture optimization~\citep{Sermpinis:Karathanasopoulos:2021, Hafiz:Broekaert:2021,Hafiz:Broekaert:2023,Yao:1999}, and impact of crises on market dynamics~\citep{Horta:Lagoa:2014,Anagnostidis:Varsakelis:2016,Smith:2012,Choi:2021}. The rest of the note is organized as follows: Section~\ref{sec:forecasting} provides the details of a day-ahead forecasting model and two crises timelines, which is followed by a detailed discussion on the co-evolution problem and different learning environments in Section~\ref{sec:coevol}. Finally, the comparative evaluation results are provided in Section~\ref{sec:res}, followed by the conclusions in Section~\ref{sec:conclusions}.

\section{Forecasting Model for Stock Indices}
\label{sec:forecasting}

\subsection{Forecasting using Technical Indicators}
\label{subsec:TImodel}

This investigation focuses on the day-ahead movement forecasting of stock indices, which can be modeled as a binary classifier,
\begin{equation}
    \label{eq:Ypred}
    y(t+1) = \begin{cases} 
            1, \quad \textit{if} \quad C(t+1)-C(t)>0,\\
            0, \quad \textit{otherwise}
    \end{cases}
\end{equation}
where, $C(t)$ gives the closing value of an index at day-$t$.

In particular, the goal is to develop a neural network based forecasting model that uses the information extracted via \emph{technical indicators}~\citep{Bustos:2020}. Such indicators provide structured input information, and, essentially, summarize the fundamental trading quantities (\textit{e.g.}, \textit{trading volume, daily open, daily close, intra-day high and low}) over a pre-fixed past period, say $[t-\tau, \ t]$ days. This study considers a total of 24 distinct technical indicators, which are evaluated with different values of $\tau$ leading to the extraction of $68$ features, see~\citep{Hafiz:Broekaert:2023} for the detailed feature extraction process. Following these steps, a labeled dataset $\mathcal{D}$ is prepared for supervised learning purposes:
\begin{equation}
    \label{eq:pattern}
    \mathcal{D} = \begin{Bmatrix} \left(x_1, x_2, \dots, x_{n_f}, y\right) ^{(k)} \ \mid \ y \in \{0,1\} \end{Bmatrix}, \qquad k = 1,2,\dots \mathcal{N}
\end{equation}
where, $x_i$ denotes the $i^{th}-$feature extracted from a particular technical indicator and $n_f$ gives the total number of features, \textit{i.e.}, $n_f=68$; $\mathcal{N}$ gives the total number of patterns in $\mathcal{D}$. For any given day$-t$, the features ($x_1,\dots, x_{n_f}$) are extracted using the trading information over the past period of $[t-\tau, t]$ days, and the corresponding day-ahead prediction label, $y$, is generated using the closing value of the next day $(t+1)$.

The next design step of effective neural forecasting models entails the selection of input features as well as appropriate neural topology. These selection issues will be discussed at length in Section~\ref{sec:coevol}. 

Further, the forecasting performance is assessed using the following three classification metrics as index movement forecasting is approached as a binary classifier: the \textit{overall accuracy}, \textit{balanced accuracy} and \textit{Matthew's Correlation Coefficient} (MCC), see~\citep{Grandini:Bagli:2020}. Notably, the earlier investigations point towards the tendency of neural networks to be biased towards a majority class~\citep{Hafiz:Broekaert:2023,UlHaq2021,Hafiz:Broekaert:2021}, \textit{e.g.}, predicting only \textit{index rises}. Such behaviors cannot be detected by the overall accuracy. Therefore, the remaining two metrics are used for supplemental evaluation. 

\begin{table*}[!t]
  \centering
  \caption{Segmentation of historical data of NASDAQ Index}
  \label{t:timeline}%
  \small
  \begin{adjustbox}{width=0.65\textwidth}
  \begin{threeparttable}
    \begin{tabular}{cccc}
    \toprule
    \textbf{Dataset} & \textbf{Role} & \makecell{\textbf{Timeline-1,}\\ \textbf{2008 Crisis}} & \makecell{\textbf{Timeline-2,}\\\makecell{\textbf{COVID-19}\\ \textbf{Pandemic}}} \\
    \midrule
    \makecell{Pre-crisis,\\ $\mathcal{D}_{pr}$} & \makecell{$^\dagger$Training data\\ for weight estimation\\ \textit{or} Test data for\\ architecture optimization} & \makecell{March, 2005\\ to December, 2006} & \makecell{January, 2017\\ to December, 2018} \\
    \midrule
    \makecell{Crisis Train,\\ $\mathcal{D}_{cr}$} & \makecell{Training data\\ for weight estimation} & \makecell{January, 2007\\ to July, 2009} & \makecell{January, 2019\\ to August, 2020} \\
    \midrule
    \makecell{Crisis Test,\\ $\mathcal{D}_{test}$} & \makecell{Test data for\\ architecture optimization} & \makecell{July, 2009\\ to April, 2010} & \makecell{August, 2020\\ to January, 2021}\\
    \midrule
    \makecell{Hold-out,\\ $\mathcal{D}_{hold}$} & \makecell{Out-of-sample data\\ for generalization evaluation} & \makecell{May, 2010\\ to May, 2011} & \makecell{January, 2021\\ to May, 2021}\\
    \bottomrule
    \end{tabular}%
    \begin{tablenotes}
      \small
      \item $^\dagger$ The pre-crisis dataset, $\mathcal{D}_{pr}$, is used in different roles depending on learning environments, see~Section~\ref{subsec:coevmulti}.
    \end{tablenotes}
  \end{threeparttable}
 \end{adjustbox}
\end{table*}%
\subsection{Historical Timelines}
\label{subsec:crisisdata}

This study considers a day-ahead forecasting of the NASDAQ index during two major market disruptions in recent history: (1) the 2008 financial crisis and (2) the COVID-19 pandemic. For investigation purposes, two timelines are being considered, where each timeline contains historical NASDAQ daily closing values a few years before and after the market disruption. Further, four distinct datasets are extracted from strictly sequestered and non-overlapping periods of each timeline. Table~\ref{t:timeline} details the role of each dataset along with the extraction time-period. The rationale here is to segment the timeline prior to and following the crisis to investigate the effects of the \emph{pre-}crisis market behavior in the forecasting period following the market disruption. For this purpose, two distinct learning environments are created where the \emph{pre-}crisis dataset, $\mathcal{D}_{pr}$, is used in an altered role, as will be discussed in Section~\ref{subsec:coevmulti}. 

Further, time-series forecasting problems are often associated with \emph{data-snooping} issues, which may lead to incorrect and inflated estimations of forecasting performance~\citep{Sullivan_EtAl_data_snooping_technical_trading_1999,Henrique_EtAl_Review_Market_prediction_snooping_2019}. To address this issue, the final evaluation of the forecasting models is determined using the hold-out dataset, $\mathcal{D}_{hold}$, which is not used in any step of the neural architecture optimization or weight estimation, and, therefore, serves as truly out-of-sample data.

\section{Multi-dataset Learning and Optimization of Neural Architectures}
\label{sec:coevol}

It is pertinent to briefly discuss the optimization of neural architectures under different learning scenarios, as the performance of neural forecasting models is critically dependent on their architectures. An efficient design of neural architecture entails optimal selection of topology (\textit{e.g.}, \emph{number} and \emph{size} of hidden layers) as well as input features (\emph{feature selection})~\citep{Yao:1999,Hafiz:Broekaert:2021,Hafiz:Broekaert:2023,Sermpinis:Karathanasopoulos:2021}. A conventional approach is to address these problems individually. However, our previous investigations in~\cite{Hafiz:Broekaert:2021,Hafiz:Broekaert:2023} demonstrated that the simultaneous selection of features and neural topology can yield better neural architectures. This is accomplished by a new \emph{co-evolution} based neural architecture selection, which considers neural architecture ($\mathcal{A}$) as a combination of the input features ($X$) and the hidden layer topology ($\mathcal{T}$), as will be discussed in Section~\ref{subsec:cobasic}. Next, Section~\ref{subsec:coevmulti} details different learning environments, which are used to test the pre-crisis data incompatibility. 

\subsection{Co-evolution: Joint optimization of features and topology}
\label{subsec:cobasic}


First, consider a classical feature selection problem which involves the identification of subset of \emph{relevant} and \emph{non-redundant} features, see~\citep{Kohavi:John:1997,Guyon:Isabelle:2003,Mehmanchi:Gomez:2021}. Let $X_{\rm full}$ denote a full set of $n_f$ number of given features then the feature selection can be formulated as:

\begin{align}
    \label{eq:fson}
     X^{\star} & = \argmin \limits_{X_i \in \Omega_F} 
     \begin{cases} 
     \mathcal{E}(X_i,\mathcal{D}_{test})\\
     \lvert X_i \rvert   \end{cases},\\
     \text{where,} \quad \Omega_F & = \Big\{ X \ \mid \ {X} \subset {X}_{\rm full} \wedge X \neq \emptyset \Big\} \qquad \text{and} \quad X_{\rm full} = \begin{Bmatrix} x_1, & x_2, & \dots, & x_{n_f} \end{Bmatrix}\nonumber
\end{align}
$X_i$ and $\mathcal{E}(X_i,\mathcal{D}_{test})$ respectively denote the $i^{th}-$subset and its classification error over a test dataset $\mathcal{D}_{test}$; $ \lvert X_i \rvert$ gives the \emph{cardinality} or number of features contained in $X_i$. $\Omega_F$ denotes the search space and it contains all possible subsets ($\approx 2^{n_f}$) of $X_{\rm full}$. The multi-objective formulation allows for the identification of a set of non-dominated feature subsets with high relevancy, by minimizing $\mathcal{E}(X)$, and low redundancy, by minimizing $\lvert X \rvert$. It is worth emphasizing that this problem is known to be an NP-Hard combinatorial problem~\citep{Guyon:Isabelle:2003}.

Next, we focus on the selection issues for topological design of hidden layers, which involve, but are not limited to: the number of hidden layers; the size (number of hidden neurons) and the activation function of each layer~\citep{Yao:1999}. Accordingly, a particular neural topology ($\mathcal{T}$) can be represented as a set of tuples, as follows:

\begin{small}
\begin{equation}
	\label{eq:nntuple}
   \mathcal{T} \leftarrow \begin{Bmatrix} (s^1,f^1), (s^2,f^2), \dots, (s^{n_\ell}, f^{n_\ell}) & \mid & s^k \in [0, s^{max}], & f^k \in \mathcal{F}, & \forall{k} \in [1,n_\ell] \end{Bmatrix}
\end{equation}
\end{small}
where, $k^{th}-$tuple $(s^k,f^k)$ gives the size and the activation function for the $k^{th}-$hidden layer; $n_\ell$ and $s^{max}$ are design inputs and respectively denote the maximum number of hidden layers and neurons; and $\mathcal{F}$ gives the pre-defined set of activation functions. Based on this formulation, the topology selection problem can be formulated as:
\begin{small}
\begin{equation}
    \label{eq:topon}
    \mathcal{T}^{\ast} = \argmin \limits_{\mathcal{T}_i \in \Omega_\mathcal{T}} 
    \begin{cases}
    \mathcal{E}(\mathcal{T}_i, \mathcal{D}_{test})\\
    \mathcal{C}(\mathcal{T}_i)
    \end{cases}, \quad \text{where,} \quad \Omega_\mathcal{T}  = \Big( [0, s^{max}] \times \mathcal{F} \Big)^{n_\ell}
\end{equation}
\end{small}
where, $\Omega_\mathcal{T}$ denotes the search space containing all possible candidate topologies; $\mathcal{E}(\cdot)$ and $\mathcal{C}(\cdot)$ determines the classification error and \emph{complexity} (to be discussed later) of the given topology.

It is worth emphasizing that usually the feature and topology selection problems, respectively given by (\ref{eq:fson}) and (\ref{eq:topon}), are pursued independently. In contrast, our previous investigations in~\cite{Hafiz:Broekaert:2021,Hafiz:Broekaert:2023} demonstrate that joint optimization of features and topology, referred to as the \emph{co-evolution} approach, can yield relatively sparse and efficacious neural architectures compared to the conventional \emph{sequential} design approaches. In essence, the co-evolution approach considers a topological perspective in which features are viewed as input neurons, which enables the integration of feature selection as a part of topological design, as follows:

\begin{small}
\begin{align}
    \label{eq:coevbasic}
    \mathcal{A}^{\ast} & = \argmin \limits_{\mathcal{A}_i \in \Omega} 
    \begin{cases}
    \mathcal{E}(\mathcal{A}_i, \mathcal{W}_{i}^{\ast}, \mathcal{D}_{test})\\
    \mathcal{C}(\mathcal{A}_i)
    \end{cases}\\
    \label{eq:W}
    \text{where,} \quad \mathcal{W}_{i}^{\ast} & = \argmin \limits_{\mathcal{W}} \ {\rm L}(\mathcal{A}_i,\mathcal{W},\mathcal{D}_{train})\\    
    \text{and} \quad \mathcal{A}_i & = \begin{Bmatrix} X_i, & \mathcal{T}_i \end{Bmatrix}, \qquad \Omega = \Omega_F \times \Omega_\mathcal{T}
\end{align}
\end{small}
$\mathcal{A}$ is a complete neural architecture which contains a subset input features, $X$, and hidden layer topology, $\mathcal{T}$; and $\Omega$ gives the joint search space of feature and topology selection problems. $\rm L(\cdot)$ and $\mathcal{D}_{train}$ respectively denote the loss function and the training dataset for the estimation of weights, $\mathcal{W}$.

Notably, the neural architecture optimization can be viewed as a \emph{bi-level} optimization problem. The upper level in~(\ref{eq:coevbasic}) can be thought of as a \emph{leader} which explores and selects architectures ($\mathcal{A}$), while the lower level in~(\ref{eq:W}) behaves as a \emph{follower} by  optimizing weights ($\mathcal{W}$) of each architecture under consideration. 

Further, it is worth noting that the estimation of architecture complexity, $\mathcal{C}(\cdot)$, in terms on nonlinear signal processing may not be trivial. This study, therefore, takes a pragmatic approach to evaluate the complexity purely in terms of topological resources which is given by,

\begin{small}
\begin{equation}
   \label{eq:complexity}
   \mathcal{C}(\mathcal{A}_i) = \frac{1}{3} \left\{ \frac{\lvert X_i \rvert}{n_f} + \frac{\Big\lvert \begin{Bmatrix} (s^k,f^k) \ \mid \ s^k \neq 0, & \forall k \in [1,n_\ell] \end{Bmatrix} \Big \rvert}{n_\ell} + \sum \limits_{k=1}^{n_\ell} \frac{s^k}{s^{max}} \right\}
\end{equation}
\end{small}
$\mathcal{C}(\cdot)$ is bounded in $[0,1]$ and it attains the maximum value of $\mathcal{C}=1$ when the architecture containing all available topological resources, \textit{e.g.}, all $n_f$ features and $n_\ell$ layers of size $s^{max}$.

\subsection{Co-evolution with Different Learning Environments}
\label{subsec:coevmulti}

In the following, we discuss the two distinct learning environments which are designed to evaluate and reconcile the effects of possibly different market dynamics following a major event with clear and enduring impact on the economy, \textit{e.g.}, \textit{2008 financial crisis}. 

The first learning environment combines \emph{pre-} ($\mathcal{D}_{pr}$) and \emph{post-}crisis ($\mathcal{D}_{cr}$) datasets, and, thereby, neglects the possible changes in the market dynamics. It is referred to as a \emph{full-learning} scenario ($\mathcal{L_F}$). The co-evolution of neural architectures under this environment is given by,

\begin{small}
\begin{align}
    \label{eq:LSF}
    \textit{co-evolution with } \mathcal{L_F}: \quad \mathcal{A}^{\ast} & = \argmin \limits_{\mathcal{A}_i \in \Omega} 
    \begin{cases}
    \mathcal{E}(\mathcal{A}_i, \mathcal{W}_{i}^{\ast}, \mathcal{D}_{test})\\
    \mathcal{C}(\mathcal{A}_i)
    \end{cases}\\
    \label{eq:W1}
    \text{where,} \quad \mathcal{W}_{i}^{\ast} & = \argmin \limits_{\mathcal{W}} \ {\rm L}(\mathcal{A}_i,\mathcal{W},\mathcal{D}_{train})\\ 
    \text{and} \quad \mathcal{D}_{train} & \leftarrow \{\mathcal{D}_{pr} \cup \mathcal{D}_{cr}\}
\end{align}
\end{small}
This scenario serves as a \emph{control} or \emph{baseline} environment for this study.

The second environment is based on the hypothesis that different market dynamics may pose learning difficulties for the estimation of weights, $\mathcal{W}$. This issue can be addressed by completely removing $\mathcal{D}_{pr}$ from the design process, \textit{e.g.}, by adjusting the size of rolling window to include only \emph{post}-crisis period. However, such an approach completely neglects any potentially correlated information in the otherwise conflicting dataset. To address this issue, the second learning environment is designed to limit the use of $\mathcal{D}_{pr}$. In particular, while $\mathcal{D}_{pr}$ serve as an additional test dataset to guide the architecture optimization process, it is not used for the weight estimation. This scenario is referred to as the \emph{split-}learning environment, $\mathcal{L}_S$, and it is given by, 

\begin{small}
\begin{align}
    \label{eq:LSE}
    \textit{co-evolution with } \mathcal{L}_S: \quad \mathcal{A}^{\ast} & = \argmin \limits_{\mathcal{A}_i \in \Omega} 
    \begin{cases}
    \mathcal{E}(\mathcal{A}_i, \mathcal{W}_{i}^{\ast}, \mathcal{D}_{test})\\
    \mathcal{C}(\mathcal{A}_i)\\
    \mathcal{E}(\mathcal{A}_i, \mathcal{W}_{i}^{\ast}, \mathcal{D}_{pr})\\
    \end{cases}\\
    \label{eq:W2}
    \text{where,} \quad \mathcal{W}_{i}^{\ast} & = \argmin \limits_{\mathcal{W}} \ {\rm L}(\mathcal{A}_i,\mathcal{W},\mathcal{D}_{train})\\ 
    \text{and} \quad \mathcal{D}_{train} & \leftarrow \mathcal{D}_{cr}
\end{align}
\end{small}
Note that in this environment weights are estimated using only the \emph{post-}crisis dataset, $\mathcal{D}_{cr}$, see~(\ref{eq:W2}). Further, the co-evolution in this environment requires optimization over three criteria; this drives the optimization process to identify sparse architectures with better classification performance in both \emph{pre-} and \emph{post-}crisis periods.

It is pertinent to highlight that the co-evolution criteria exhibit a certain degree of conflict regardless of the learning environment. For instance, a reduction in the complexity of an over-fitted architecture is expected to be accompanied by the reduction in classification errors. On the other hand, excessively sparse architectures, characterized by substantial reduction in the complexity, may suffer from high bias errors. A unique optimal architecture, $A^\star$, which minimizes all co-evolution criteria, thus, may not exist. The goal of co-evolution is, therefore, to identify a set of \emph{non-dominated} architectures, Pareto set $\Gamma^\star$, which represent distinct compromise over different criteria, as follows:
\begin{align}
    \Gamma^\star = \Big\{ \mathcal{A} \in \Omega \ \mid \ \nexists{\mathcal{A}_j} \in \Omega : \mathcal{A}_j \preceq \mathcal{A}  \Big\}
\end{align}
where, $\mathcal{A}_j\preceq \mathcal{A}_k$ denotes $\mathcal{A}_j$ \emph{dominates} $\mathcal{A}_k$. 
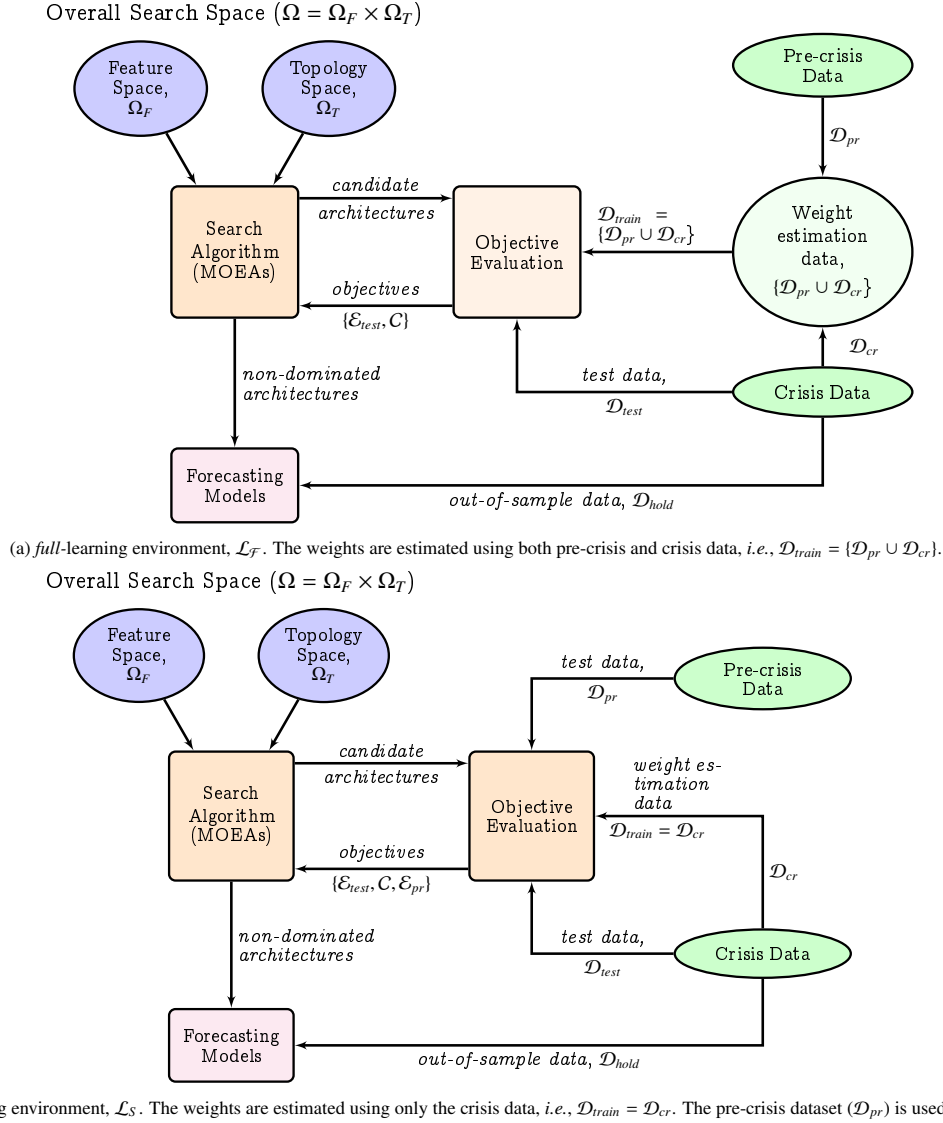
\begin{figure}[!t]
\centering
\tikzstyle{block} = [rectangle, draw, fill=white, 
        text width=7em, text centered, rounded corners, minimum height=4.5em,line width=0.05cm]
\tikzstyle{block2} = [rectangle, draw, fill=white, 
    text width=7em, text centered, rounded corners, minimum height=8em,line width=0.05cm]
\tikzstyle{line} = [draw, -latex']
\tikzstyle{cloud} = [draw, ellipse,fill=white, node distance=4.8cm, minimum height=3em, line width=0.05cm,text width=7em,text centered]
\tikzstyle{cloud2} = [draw, ellipse,fill=white, node distance=4.8cm, minimum height=3em, line width=0.05cm,text width=5em,text centered]
\begin{subfigure}{\textwidth}
\centering
\begin{adjustbox}{max width=0.7\textwidth}
  \centering
    \begin{tikzpicture}[auto]
        \centering
       \node [cloud2, fill=blue!20] at (-2,6.5) (FS) {\large Feature Space, $\Omega_F$};
       \node [cloud2, fill=blue!20] at (2,6.5) (TS) {\large Topology Space, $\Omega_T$};
       \node[fit=(FS)(TS),label={[node font=\Large,black]above:Overall Search Space ($\Omega= \Omega_F \times \Omega_T$)}](F){};
       \node [block2, fill=orange!20] at (0,3) (search) {\large Search\\Algorithm\\ (MOEAs)}; 
       \node [block2, fill=orange!10] at (6,3) (OE) {\large Objective Evaluation};
       \node[cloud,fill=green!20] at (12.5,0) (data) {\large Crisis Data}; 
       \node[cloud,fill=green!20] at (12.5,7) (predata) {\large Pre-crisis Data}; 
       \node[cloud,fill=green!5] at (12.5,3) (trdata) {\large Weight\\ estimation data,\\[1ex] $\{\mathcal{D}_{pr} \cup \mathcal{D}_{cr}$\}}; %
       \node[block,fill=magenta!10] at (0,-2) (final){\large Forecasting Models};    
        \path [line,ultra thick] (FS) -- (search);
        \path [line,ultra thick] (TS) -- (search); 
         \path [line,ultra thick] (search) --node[right,text width=3.5cm]{\large \emph{non-dominated architectures}} (final);   
        \path [line, ultra thick] (search.40) --node [above]{\large\textit{candidate}} node [below]{\large \textit{architectures}} (OE.140); 
        \path [line, ultra thick] (OE.220) --node [above]{\large\textit{objectives}} node[below] {\large $\{\mathcal{E}_{test}, \mathcal{C} \}$}(search.320); 
        \path [line,ultra thick] (predata.270) --node [right]{\large $\mathcal{D}_{pr}$} (trdata.90);
        \path [line,ultra thick] (data.90) --node [right]{\large $ \quad \mathcal{D}_{cr}$} (trdata.270);
        \path [line,ultra thick] (trdata.180) -- node[above,text width=2.5cm] {\large $\mathcal{D}_{train} = \{\mathcal{D}_{pr} \cup \mathcal{D}_{cr}$\}} (OE.0);
        \path [line,ultra thick] (data.180) -|node [below,near start]{\large $\mathcal{D}_{test}$}  node[above,near start] {\large \textit{test data,}} (OE.270);
       \path [line,ultra thick] (data.270) |-node [below, near end]{\large \emph{out-of-sample data}, \large $\mathcal{D}_{hold}$} node[below, near end] {} (final.0);
    \end{tikzpicture}  
\end{adjustbox}
\caption{\emph{full-}learning environment, $\mathcal{L_F}$. The weights are estimated using both pre-crisis and crisis data, \textit{i.e.}, $\mathcal{D}_{train} = \{ \mathcal{D}_{pr} \cup \mathcal{D}_{cr}\}$.}
\label{f:LSF}
\end{subfigure}\\
\begin{subfigure}{\textwidth}
  \centering
  \begin{adjustbox}{max width=0.7\textwidth}
    \begin{tikzpicture}[auto]
        \centering
       \node [cloud2, fill=blue!20] at (-2,6.5) (FS) {\large Feature Space, $\Omega_F$};
       \node [cloud2, fill=blue!20] at (2,6.5) (TS) {\large Topology Space, $\Omega_T$};
       \node[fit=(FS)(TS),label={[node font=\Large,black]above:Overall Search Space ($\Omega= \Omega_F \times \Omega_T$)}](F){};
       \node [block2, fill=orange!20] at (0,3) (search) {\large Search\\Algorithm\\ (MOEAs)}; 
       \node [block2, fill=orange!20] at (6.5,3) (OE) {\large Objective Evaluation};
       \node[cloud,fill=green!20] at (11.5,0) (data) {\large Crisis Data}; 
       \node[cloud,fill=green!20] at (11.5,6) (predata) {\large Pre-crisis Data}; 
       \node[block,fill=magenta!10] at (0,-2) (final){\large Forecasting Models};  
        \path [line,ultra thick] (FS) -- (search);
        \path [line,ultra thick] (TS) -- (search); 
         \path [line,ultra thick] (search) --node[right,text width=3.5cm]{\large \emph{non-dominated architectures}} (final);
        \path [line, ultra thick] (search.40) --node [above]{\large\textit{candidate}} node [below]{\large \textit{architectures}} (OE.140); 
        \path [line, ultra thick] (OE.220) --node [above]{\large\textit{objectives}} node[below] {\large $\{\mathcal{E}_{test}, \mathcal{C}, \mathcal{E}_{pr} \}$}(search.320);     
        \path [line,ultra thick] (predata.180) -|node [below,near start]{\large $\mathcal{D}_{pr}$} node [above,near start]{\large \textit{test data,}}(OE.90);  
        \path [line,ultra thick] (data.180) -|node [below,near start]{\large $\mathcal{D}_{test}$}  node[above,near start] {\large \textit{test data,}} (OE.270);  
        \path [line,ultra thick] (data.90) |-node [right,near start,text width=3cm]{\large $\mathcal{D}_{cr}$} node[above, near end, text width=2cm]{\large \emph{weight estimation data}} node[below, near end, text width=3cm]{\large $\mathcal{D}_{train}=\mathcal{D}_{cr}$}(OE.0); 
       \path [line,ultra thick] (data.270) |-node [below, near end]{\large \emph{out-of-sample data}, \large $\mathcal{D}_{hold}$} node[below, near end] {} (final.0);
    \end{tikzpicture}  
    \end{adjustbox}
    \caption{\emph{split-}learning environment, $\mathcal{L}_S$. The weights are estimated using only the crisis data, \textit{i.e.}, $\mathcal{D}_{train} = \mathcal{D}_{cr}$. The pre-crisis dataset ($\mathcal{D}_{pr}$) is used as a test dataset.}
    \label{f:LSE}
\end{subfigure}
\caption{Multi-objective co-evolution framework for simultaneous optimization of features and neural topology under different multi-dataset learning scenarios. $\mathcal{D}_{pr}$ denotes the learning dataset derived from the pre-crisis period. $\mathcal{D}_{cr}$, $\mathcal{D}_{test}$, and $\mathcal{D}_{hold}$ respectively denote the training, testing and \emph{out-of-sample} datasets derived from the crisis time-period. Note that $\mathcal{D}_{hold}$ is not used in any step of model development and strictly sequestered to evaluate the out-of-sample performance of the final forecasting models. We refer to~\cite{Hafiz:Broekaert:2023} for a detailed treatment on this search framework and pseudo-codes.}
\label{f:search}
\end{figure}

Further, it is crucial to emphasize that the neural architecture search under any learning environment represents a NP-Hard problem, which necessitates a proficient search framework. To this end, the search framework proposed in ~\cite{Hafiz:Broekaert:2023} is used, which essentially consists of a Multi-Objective Evolutionary Algorithms (MOEA) with a posteriori decision making tool. 

Fig.~\ref{f:search} outlines the co-evolution search framework for both the learning environments. The key component of this framework is the MOEA, which explores the search space of co-evolution problem ($\Omega$) to generate candidate architectures ($\mathcal{A}$). In essence, each search iteration of MOEA involves evolution, and, subsequent, evaluation of candidate architectures. To this end, each architecture under consideration is evaluated over either two or three objectives depending on the learning environment, see `Objective Evaluation' in Fig.~\ref{f:LSF}~and~\ref{f:LSE}, as follows: The complexity of architectures, $\mathcal{C}(\cdot)$, is determined using~(\ref{eq:complexity}). Next, the classification performance of the architectures is determined in terms of \emph{balanced error}, see~\cite{Grandini:Bagli:2020}. The rationale here is to address the tendency of neural networks to be biased towards a particular class, say predicting only \emph{index rises}, as highlighted in the earlier investigations~\citep{Hafiz:Broekaert:2023,UlHaq2021,Hafiz:Broekaert:2021}. Accordingly, $\mathcal{E}(\cdot)$ denotes a \emph{balanced error} over the given dataset in~(\ref{eq:LSF})~and~(\ref{eq:LSE}). Further, for the rest of this article we denote $\mathcal{E}(\mathcal{A}_i, \mathcal{W}_{i}^{\ast}, \mathcal{D}_{test})$ and $\mathcal{E}(\mathcal{A}_i, \mathcal{W}_{i}^{\ast}, \mathcal{D}_{pr})$ respectively as $\mathcal{E}_{test}(\mathcal{A}_i)$ and $\mathcal{E}_{pr}(\mathcal{A}_i)$ for notational brevity.

Note that a set non-dominated architectures identified at the end of iterative search process of MOEA (say $\gamma$) represents an approximation of true Pareto front, $\Gamma^\star$, owing to deceptive, noisy and multi-modal search space of the co-evolution problem~\citep{Yao:1999,Guyon:Isabelle:2003}. An effective multi-objective search philosophy is, therefore, essential. Consequently, the following two distinct MOEAs are selected to accommodate distinct search philosophies: NSGA-II~\citep{Deb:Pratap:2002} which is augmented with \emph{non-geometric} crossover operator~\citep{Ishibuchi:Tsukamoto:2010}; and External Archive Guided Decomposition based MOEA (EAGD)~\citep{Cai:Li:2015}. A detailed discussion on the search framework and algorithms is omitted here for sake of brevity; we refer to~\cite{Hafiz:Broekaert:2023} for comprehensive details of the search framework and to~\citep{Cai:Li:2015,Deb:Pratap:2002,Hafiz:Swain:MOEA:2020} for a detailed treatment on dominance-decomposition search philosophies of MOEAs.

\section{Results}
\label{sec:res}

\subsection{Experimental Set-up}
\label{subsec:setup}


We begin by considering two major stock market disruptions in recent history to investigate a potential shift in market dynamics. These disruptions are contained in two `Timelines' being investigated: Timeline-1 (2008 crisis) and Timeline-2 (COVID-19 Pandemic), see Section~\ref{subsec:crisisdata}. Next, neural architectures are optimized for each timeline under two different learning environments, where the roles assigned to the pre-crisis dataset differ, see $\mathcal{L}_\mathcal{F}$ and $\mathcal{L}_\mathcal{S}$ in Section~\ref{subsec:coevmulti}. The goal here is to compare a set of \emph{non-dominated} neural forecasting models which are selected with the altered roles of the \emph{pre-}crisis dataset, $\mathcal{D}_{pr}$. Finally, in each learning environment, two distinct MOEAs, non-geometric NSGA-II~\citep{Deb:Pratap:2002,Ishibuchi:Tsukamoto:2010} and EAGD~\citep{Cai:Li:2015}, are used to identify non-dominated architectures. 

To summarize, a total of four search scenarios ($2$ \textit{learning environments} $\times$ $2$ \textit{MOEAs}) are being investigated for each timeline, as shown in Table~\ref{t:search}. Each search scenario is designed as follows to ensure an objective comparative evaluation: A total of $68$ input features are extracted from 24 technical indicators and basic trading information, \textit{i.e.}, $n_f=68$ and the corresponding feature space, $\Omega_F = 2^{68}$. Each architecture can have a maximum of two hidden layers where the maximum size of each layer is limited to $128$, \textit{i.e.}, $n_\ell=2$ and $s^{max}=128$. Further, the activation function for a particular hidden layer can be selected from $\mathcal{F}=\{tansig, \ logsig\}$, which gives the topology space of $\Omega_\mathcal{T} = \Big( [0, s^{max}] \times \mathcal{F} \Big)^{n_\ell}$. A total of $40$ independent runs of MOEAs are carried out in this search space ($\Omega_F \times \Omega_\mathcal{T}$), where each run is set to terminate after $300$ search iterations. The approximate Pareto set identified at the end of each run is stored and used for the subsequent comparative analysis. The search parameters of MOEAs are set empirically by considering the recommendations of the earlier investigations~\citep{Cai:Li:2015,Ishibuchi:Tsukamoto:2010,Hafiz:Swain:MOEA:2020}, as follows:
\begin{itemize}
    \item NSGA-II with non-geometric crossover: \textit{population size}$\rightarrow50$; \textit{crossover rate}$\rightarrow0.9$; \textit{probability of non-geometric crossover}$\rightarrow0.8$; \textit{probability of bit-flip}$\rightarrow1/n$; \textit{mutation rate}$\rightarrow1/n$, where $n$ is the total number of search variables.
    \smallskip
    \item EAGD: \textit{population size}$\rightarrow50$; \textit{crossover rate}$\rightarrow1.0$; \textit{mutation rate}$\rightarrow1/n$; \textit{learning generations:}$\rightarrow8$; \textit{neighbors:}$\rightarrow10\%$ of population size; a set uniformly distributed weights for decomposition sub-problems are generated as per~\cite{Das:Dennis:1998}. 
\end{itemize}

Further, it is pertinent to note that the number of co-evolution criteria depends on the learning environment, see~(\ref{eq:LSF})~and~(\ref{eq:LSE}); the co-evolution is a three-criteria optimization problem under $\mathcal{L_S}$, where the additional criterion is to improve the classification performance over the \emph{pre-}crisis period, $\mathcal{D}_{pr}$, see~(\ref{eq:LSE}). This difference in the number of criteria is addressed by considering the classification performance over the \emph{out-of-sample} dataset $\mathcal{D}_{hold}$ for the comparative evaluation, \textit{i.e.}, \emph{balanced error} over $\mathcal{D}_{hold}$, denoted by $\mathcal{E}_{hold}(\cdot)$. Such evaluation not only allows for the comparison of learning environments over the same criteria, $\mathcal{E}_{hold}(\cdot)$ and $\mathcal{C}(\cdot)$, but it also minimizes the possible effects of \emph{data-snooping}~\citep{Sullivan_EtAl_data_snooping_technical_trading_1999,Henrique_EtAl_Review_Market_prediction_snooping_2019}, as $\mathcal{D}_{hold}$ is not used in any step of the architecture optimization or weight estimation processes, see Fig.~\ref{f:search}.
\begin{table*}[!t]
  \centering
  \caption{Search scenarios associated with each timeline}
  \label{t:search}%
  \small
  \begin{adjustbox}{width=0.45\textwidth}
  \begin{threeparttable}
    \begin{tabular}{lcc}
    \toprule
    \textbf{Scenario} & \makecell{\textbf{Learning}\\\textbf{Environment}\\(\textbf{Role of} \boldmath$\mathcal{D}_{pr}$)} & \makecell{\textbf{MOEA}\\\textbf{Philosophy}} \\
    \midrule
    $\mathcal{L}_\mathcal{F}$ + NSGA-II & \multirow{2}[2]{*}{\makecell{Training\\(\textit{weight estimation})}} & \textit{dominance} \\[1ex]
    $\mathcal{L}_\mathcal{F}$ + EAGD &       & \makecell{\textit{decomposition}\\+ \textit{dominance}} \\[1ex]
    \midrule
    $\mathcal{L}_\mathcal{S}$ + NSGA-II & \multirow{2}[2]{*}{\makecell{Testing\\(\textit{architecture search}\\\textit{criterion})}} & \textit{dominance} \\[1ex]
    $\mathcal{L}_\mathcal{S}$ + EAGD &       & \makecell{\textit{decomposition}\\+ \textit{dominance}} \\
    \bottomrule
    \end{tabular}%
  \end{threeparttable}
 \end{adjustbox}
\end{table*}%

\subsection{Comparative Evaluation of Learning Environments}
\label{subsec:compareeval}

The co-evolution problem is solved for each timeline under four search scenarios (see Table~\ref{t:search}) to compare the \emph{full} and \emph{split} learning environments. The goal here is to identify whether the learning environments, $\mathcal{L_F}$ and $\mathcal{L}_S$, significantly impact the identified Pareto sets. 

A total of 40 independent runs of MOEAs, NSGA-II or EAGD, are carried out in each search scenario to accommodate the stochastic nature of evolutionary algorithms. The approximate Pareto set identified in each run, say $\gamma$, is stored and used for the subsequent comparative analysis, \textit{e.g.},
\begin{align}
    \Gamma_{FN} = \begin{Bmatrix} \gamma_{FN,1}, & \gamma_{FN,2}, & \dots  \end{Bmatrix}
\end{align}
where, $\gamma_{FN,i}$ denotes the approximate Pareto set (APS) identified under the full learning environment ($\mathcal{L_F}$) and in the $i^{th}-$run of the NSGA-II algorithm; and $\Gamma_{FN}$ denotes the set of different APS  which are identified over multiple runs of NSGA-II. 

Further, this study considers the hyper-volume (HV) indicator metric~\citep{Fonseca:Paquete:2006,Ishibuchi:Imada:2017} to compare APS identified under different learning environments. The hyper-volume, in essence, gauges the area in criteria space which is dominated by the APS under consideration and bounded from above by a reference point. The HV indicator measures the ratio of hyper-volumes of the given APS and the true Pareto set; a \emph{higher} value of this metric is, therefore, desirable. Further, given that the true Pareto set of the co-evolution problem is not known, it is approximated by using the union of all APS identified over 40 independent runs of MOEA and under all search scenarios, as follows:
\begin{align}
\Gamma_{union} & = \{\Gamma_{FN} \cup \Gamma_{FE} \cup \Gamma_{SN} \cup \Gamma_{SE}\}\\
\Gamma^\ast & = \{\mathcal{A} \in \Gamma_{union} \, \mid \, \nexists \mathcal{A}_i \in \Gamma_{union} : \mathcal{A}_i \preceq \mathcal{A} \}
\end{align}
where, $\Gamma^\ast$ gives the estimated true Pareto set; $\Gamma_{FN}$ and $\Gamma_{SN}$ denotes the set of APS identified over multiple runs of NSGA-II under $\mathcal{L_F}$ and $\mathcal{L}_S$ environments, respectively; similarly, $\Gamma_{FE}$ and $\Gamma_{SE}$ respectively denotes the set of APS identified over multiple runs of EAGD under $\mathcal{L_F}$ and $\mathcal{L}_S$.

This study considers the dimension sweep algorithm by~\cite{Fonseca:Paquete:2006} to calculate the HV indicator. We follow the recommendations of~\cite{Ishibuchi:Imada:2017} to reduce the sensitivity to the reference point specification in the hyper-volume calculations. 

The mean and standard deviation of the HV indicators over 40 independent runs of MOEAs for each scenario are listed in Table~\ref{t:HV2008} (\textit{for Timeline-1, 2008 crisis}) and in Table~\ref{t:HVcovid} (\textit{for Timeline-2, COVID-19 pandemic}). It is clear that for both timelines, the search scenarios with the $\mathcal{L}_S$ learning environment yield relatively higher mean values of the HV indicator. 

\begin{table*}[!t]
  \centering
  \footnotesize
   \caption{HV Indicator and statistical comparison over different search scenario: Timeline-1 (2007-2008 Crisis)}
  \label{t:HV2008}%
  \small
  \begin{adjustbox}{width=0.55\textwidth}
  \begin{threeparttable}
    \begin{tabular}{lccc}
    \toprule
    \textbf{Scenario} & \makecell{\textbf{Hyper-volume}\\ \textbf{Indicator}\\\textbf{(Mean $\pm$ SD)}} & \boldmath $p-$\textbf{value} & \makecell{ $^\dagger$\textbf{Adjusted}\\\boldmath$p-$\textbf{value}\\ \textbf{(APV)}} \\[0.5ex]
    \midrule
     $\mathcal{L_F}$ + NSGA-II & 0.9296 $\pm$ 1.52E-02 & 5.09E-7 & 2.50E-02$^{\ast\ast}$\\ [0.5ex]
    $\mathcal{L_F}$ + EAGD & 0.9218 $\pm$ 1.82E-02 & 1.34E-9 & 1.67E-02$^{\ast\ast}$\\[0.5ex]
    $\mathcal{L}_S$ + NSGA-II & 0.9497 $\pm$ 1.45E-02 & 4.90E-01 & 5.00E-02$^{\ast\ast}$\\[0.5ex]
    $\mathcal{L}_S$ + EAGD$^\ddagger$ & \textbf{0.9532} $\pm$ 3.11E-02 & - & -\\
    \bottomrule
    \end{tabular}%
    \begin{tablenotes}
      \footnotesize
      \item $^\ddagger$ The $p-$value of Friedman test is $p=5.98E-11$. $\mathcal{L}_S$ + EAGD scenario obtained the best Friedman ranking and serves as the \emph{control} for the multiple comparisons in the posthoc analysis
      \item  $^\dagger$  null hypothesis states that a particular approach is better than the control scenario: $\mathcal{L}_S$ + EAGD. 
      \item $^\dagger$ Hommel's posthoc procedure suggests that all null-hypotheses with APV$\le0.05$ should be rejected at $\alpha=0.05$ level (95\% confidence interval)
      \item $^\dagger$ $^{\ast\ast}$ denotes that null-hypothesis is rejected
    \end{tablenotes}
  \end{threeparttable}
 \end{adjustbox}
\end{table*}%
\begin{table*}[!t]
  \centering
  \footnotesize
  \caption{HV Indicator and statistical comparison over different search scenario: Timeline-2 (COVID-19 Pandemic)}
  \label{t:HVcovid}%
  \small
  \begin{adjustbox}{width=0.55\textwidth}
  \begin{threeparttable}
    \begin{tabular}{lccc}
    \toprule
    \textbf{Scenario} & \makecell{\textbf{Hyper-volume}\\ \textbf{Indicator}\\\textbf{(Mean $\pm$ SD)}} & \boldmath $p-$\textbf{value} & \makecell{ $^\dagger$\textbf{Adjusted}\\\boldmath$p-$\textbf{value}\\\textbf{(APV)}} \\[0.5ex]
    \midrule	
     $\mathcal{L_F}$ + NSGA-II & 0.9001 $\pm$ 1.35E-02 & 4.51E-10 & 2.50E-02$^{\ast\ast}$\\ [0.5ex]
    $\mathcal{L_F}$ + EAGD & 0.8941 $\pm$ 1.56E-02 & 9.49E-14 & 1.67E-02$^{\ast\ast}$\\[0.5ex]
    $\mathcal{L}_S$ + EAGD & 0.9186 $\pm$ 2.46E-02 & 9.98E-4 & 5.00E-02$^{\ast\ast}$\\ [0.5ex]
    $\mathcal{L}_S$ + NSGA-II$^\ddagger$ & \textbf{0.9443} $\pm$ 2.34E-02 & - & -\\
    \bottomrule
    \end{tabular}%
    \begin{tablenotes}
      \footnotesize
      \item  $^\ddagger$ The $p-$value of Friedman test is $p=3.71E-11$. $\mathcal{L}_S$ + NSGA-II scenario obtained the best Friedman ranking and serves as the \emph{control} for the multiple comparisons in the posthoc analysis
      \item $^\dagger$  null hypothesis states that a particular approach is better than the control scenario: $\mathcal{L}_S$ + NSGA-II.
      \item $^\dagger$ Hommel's posthoc procedure suggests that all null-hypotheses should be rejected at $\alpha=0.05$ level (95\% confidence interval) 
      \item $^\dagger$ $^{\ast\ast}$ denotes that null-hypothesis is rejected
    \end{tablenotes}
  \end{threeparttable}
 \end{adjustbox}
\end{table*}%

\begin{figure*}[!t]
\centering

\begin{subfigure}{.48\textwidth}
  \centering
  \includegraphics[width=\textwidth]{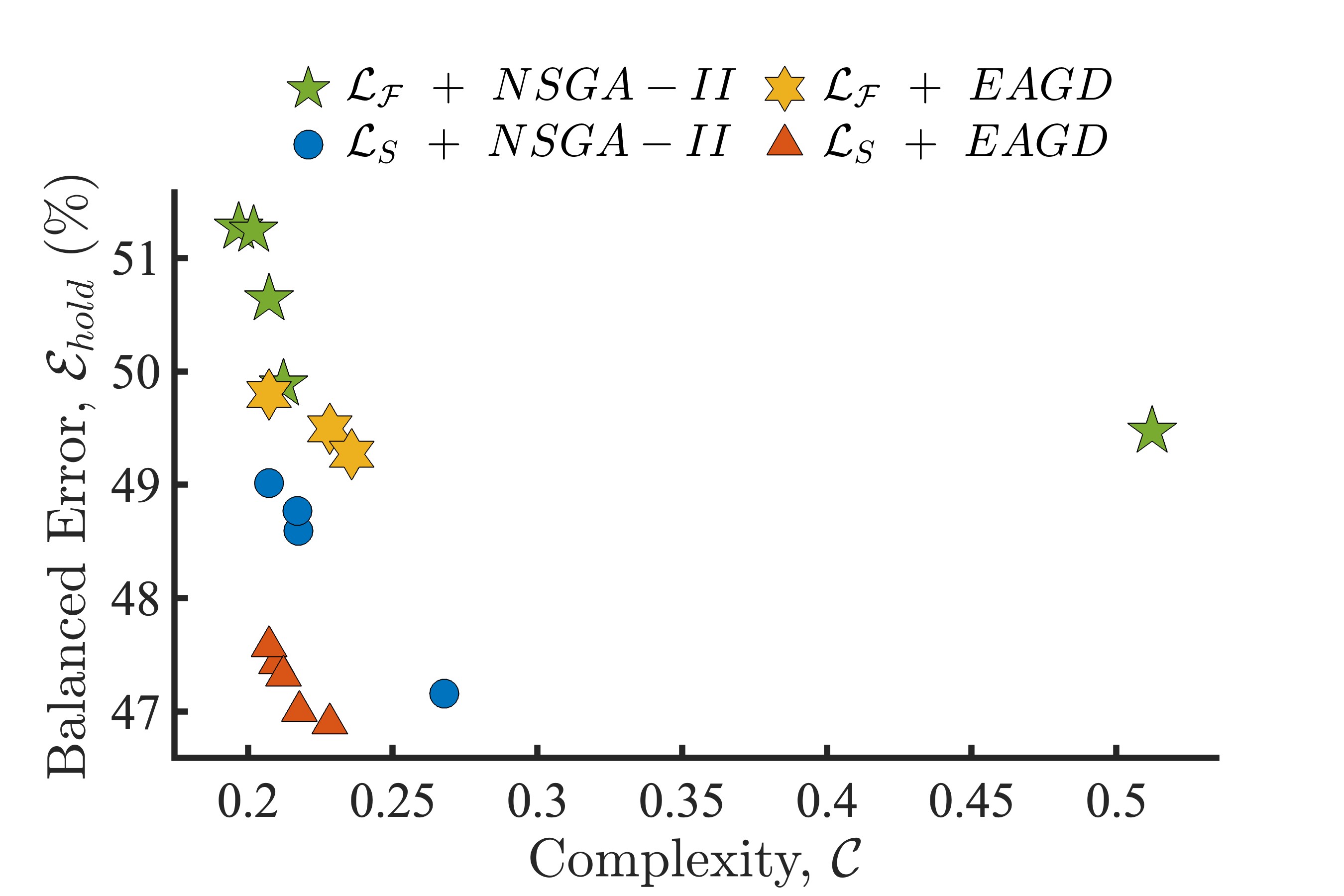}
  \caption{Timeline - 1: 2008 Financial Crisis}
  \label{f:median2008}
\end{subfigure}
\begin{subfigure}{.48\textwidth}
  \centering
  \includegraphics[width=\textwidth]{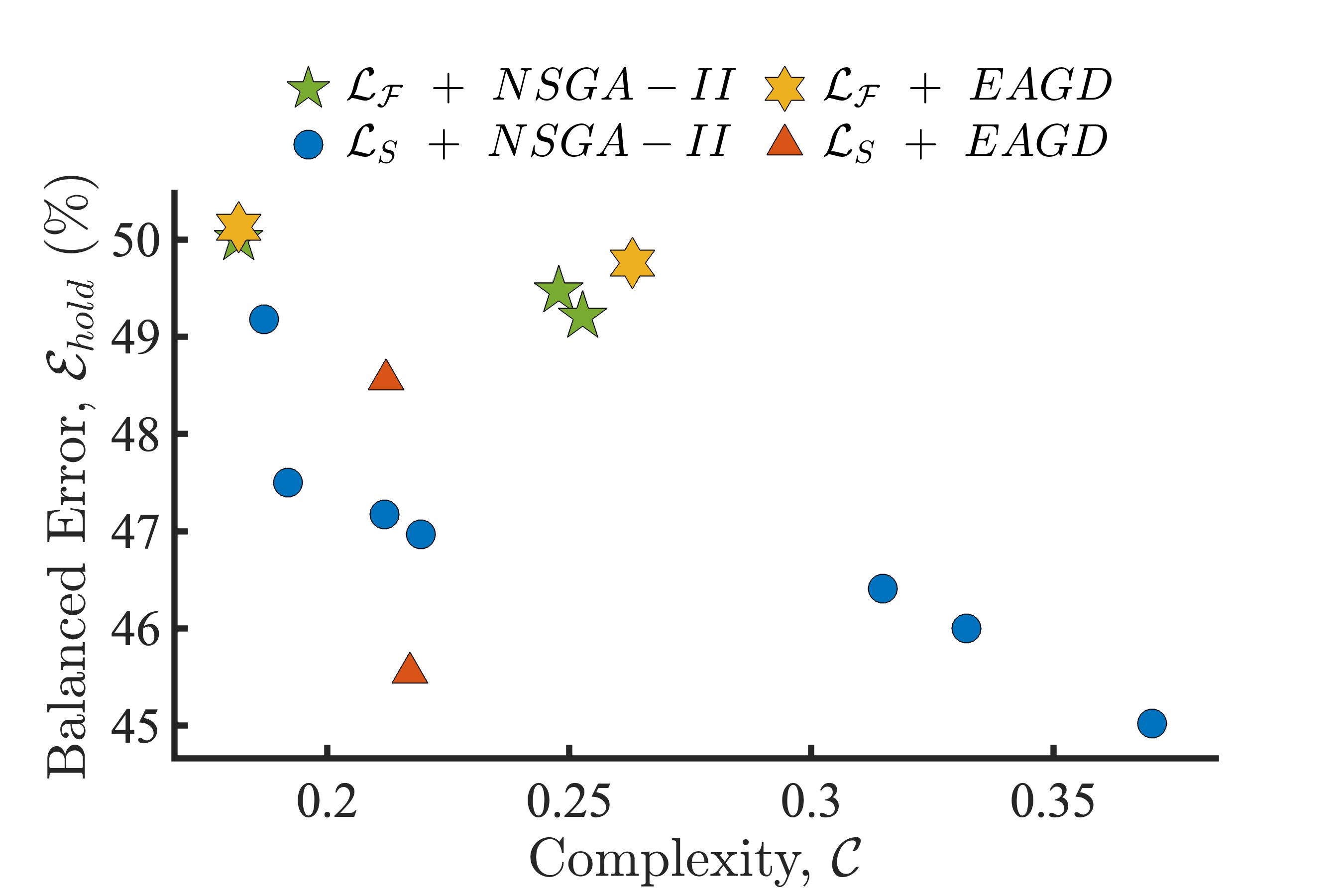}
  \caption{Timeline - 2: COVID -19 Pandemic}
  \label{f:medianCOVID}
\end{subfigure}
\caption{Median front over 40 independent runs of MOEAs over both timelines}
\label{f:median}
\end{figure*}

Further, a \emph{median front} is determined for each search scenario to visualize these results. The median front is essentially the Pareto front corresponding to the APS with the median value of the HV indicator. The median fronts corresponding to both timelines are shown in Fig.~\ref{f:median}. The comparison of the median fronts clearly highlights the dominance of the $\mathcal{L}_S$ learning environment in both timelines. Further, the results also highlight the effects of the search philosophies of MOEAs. For instance, the hybrid decomposition-dominance philosophy of EAGD~\citep{Cai:Li:2015} is found to be relatively more effective in the 2008 crisis timeline, see Fig.~\ref{f:median2008}. On the other hand, NSGA-II with the non-geometric operator could yield relatively wider fronts in the COVID-19 timeline, especially with the $\mathcal{L}_S$ environment, see Fig.~\ref{f:medianCOVID}. 

While the median fronts in Fig.~\ref{f:median} provide a good visual comparison, further analysis is essential to compare the results over multiple independent runs of MOEAs. To this end, non-parametric statistical tests are carried out following the guidelines in~\citep{Derrac:Salvador:2011,Garcia:Salvador:2010} to check the significance of the results obtained over 40 independent runs. In particular, the Friedman test is carried out first to test the null-hypothesis that specifies equality of median HV indicator value under all search scenarios. This hypothesis can safely be rejected for both timelines with the following $p-$values: $5.98\times10^{-11}$ (\textit{Timeline-1}), $3.71\times10^{-11}$ (\textit{Timeline-2}). Further, the best search scenario according to the average ranking is found to be $\mathcal{L}_S$ + EAGD (Timeline-1) and $\mathcal{L}_S$ + NSGA-II (Timeline-2). Accordingly, these scenarios serve as the \emph{control} scenario in the subsequent post hoc analysis.

The second stage of statistical comparison involves multiple comparisons of the control with the other search scenarios to test the null-hypothesis that the scenario being compared is better than the control. The Hommel's post-hoc procedure is followed for this purpose. The adjusted $p-$values (APV) following this procedure are shown in Table~\ref{t:HV2008}~and~\ref{t:HVcovid}, which indicate that null-hypotheses can be rejected at a $\alpha=0.05$ significance level. These results clearly underline that the split learning environment $\mathcal{L}_S$ can evolve better APS of the co-evolution problem for both timelines. Given that the neural architecture are evolved under a similar set-up in both learning environments, the significant difference in results can be attributed to the altered role of the \emph{pre}-crisis dataset, $\mathcal{D}_{pr}$. 

\begin{table*}[!t]
  \centering
  \footnotesize
   \caption{Selected architecture}
  \label{t:selarch}%
  \begin{adjustbox}{width=0.75\textwidth}
  \begin{threeparttable}

    \begin{tabular}{cccccc}
    \toprule
    \textbf{Scenario} & \makecell{\textbf{Selected}\\ \textbf{Features}\\ \boldmath$\lvert X \rvert$} & \makecell{\textbf{Hidden}\\\textbf{Layer-1}\\ \boldmath$(s^1,f^1)$} & \makecell{\textbf{Hidden}\\\textbf{Layer-2}\\ \boldmath$(s^2,f^2)$} & \makecell{\textbf{Complexity,}\\\boldmath$\mathcal{C}$} & \makecell{$^{\dagger\dagger}$\textbf{Classification}\\\textbf{Performance}\\\textbf{on} \boldmath$\mathcal{D}_{hold}$} \\

    \midrule
    Timeline-1 & \multirow{2}{*}{14} & 96    & -     & \multirow{2}{*}{0.4873} & 58.35 \\[1ex]
    $\mathcal{L}_S$ + EAGD &       & \textit{tansig} & -     &       & (0.20) \\

    \midrule
    Timeline-2  & \multirow{2}{*}{5} & 32    & -     & \multirow{2}{*}{0.2752} & 56.32 \\[1ex]
      $\mathcal{L}_S$ + NSGA-II &       & \textit{tansig} & -     &       & (0.10) \\
      
    \bottomrule
    \end{tabular}%
    
    \begin{tablenotes}
      \footnotesize
      \item $^{\dagger\dagger}$ - forecasting performance in terms of overall accuracy/hit-rate (in percentage) and Matthews Correlation Coefficient (MCC); a higher value is desirable for both metrics. The values of MCC are shown inside parentheses.
    \end{tablenotes}
  \end{threeparttable}
 \end{adjustbox}
\end{table*}%

Finally, an illustrative neural architecture is selected from the set of identified APS over multiple runs of MOEAs to demonstrate the generalization performance. For this purpose, the \emph{a posteriori} architecture selection approach suggested in~\citep{Hafiz:Broekaert:2023,Hafiz:Swain:MOEA:2020} is followed to select architectures from the best search scenario of each timeline, \textit{i.e.}, $\mathcal{L}_S$ + EAGD (Timeline-1, 2008 crisis) and $\mathcal{L}_S$ + NSGA-II (Timeline-2, COVID-19 pandemic). Table~\ref{t:selarch} shows the selected architectures and the corresponding average classification performance on the \emph{out-of-sample} dataset $\mathcal{D}_{hold}$ over 50 independent training-testing cycles. Note that overall classification accuracy is analogous to the financial metric \emph{hit-rate}, which also measures the ratio of correct to overall predictions~\citep{Shynkevich:McGinnity:2017}. 


\section{Conclusions}
\label{sec:conclusions}

The effects of a potential shift in stock trading behavior following a major market disruption was investigated in the context of the fundamental problem of neural architecture design for forecasting. An empirical test of the pre-crisis data incompatibility was carried out through the comparative evaluation on two major market crisis periods, the 2008 crisis and the COVID-19 pandemic. The results show a statistically significant improvement in the hypervolume indicator metric when the role of the pre-crisis data is limited, which indicates a better convergence to the true Pareto set of the co-evolution problem, \textit{i.e.}, neural architectures with both lower generalization error and complexity are identified when the pre-crisis dataset is limited only to the architecture optimization and not used in the weight estimation ($\mathcal{L}_S$). 

The  observations in our present note demonstrate the occurrence of learning inconsistencies in \emph{pre-} and \emph{post-}crisis market data and, thereby, highlight the need for a reconciliation framework, like $\mathcal{L}_S$, to address the evolving market behavior. Note that our present results are in agreement with earlier investigations, where the need for such a reconciliation method was also emphasized, \textit{e.g.}, \cite{Gerlein:McGinnity:2016} demonstrated that adjusting the length of the sliding window for model retraining can have a significant impact on forecasting performance.

\bibliographystyle{elsarticle-num}

\end{document}